\documentstyle[12pt]{article}
\def\simless{\mathbin{\lower 1pt\hbox
   {$\spose{\raise 5pt\hbox{$\char'074$}}\char'430$}}}
\def\simgreat{\mathbin{\lower 1pt\h   {$\spose{\raise 5pt\hbox{$\char'076$}}\char'430$}}}
\def\simgreat{\gapp}
\def\simless{\lapp}
\def\lapp{\mathbin{\raise2pt \hbox{$<$} \hskip-9pt \lower4pt \hbox{$\sim$}}}
\def\gapp{\mathbin{\raise2pt \hbox{$>$} \hskip-9pt \lower4pt \hbox{$\sim$}}}
\newcommand{\rfnce}{\par \noindent \hangindent 15pt{}}

\newcommand{\be}{\begin{equation}}
\newcommand{\ee}{\end{equation}}

\newcommand{\bea}{\begin{eqnarray}}
\newcommand{\eea}{\end{eqnarray}}

\newcommand{\al}{\alpha}

\newcommand{\dl}{\delta}

\newcommand{\Th}{\Theta}

\newcommand{\lm}{\lambda}
\newcommand{\ks}{\xi}

\newcommand{\Ph}{\Phi}

\newcommand{\nn}{\nonumber}

\begin{document}

\title
{\large \bf SEMICLASSICAL QUADRATIC GRAVITY REVISITED \footnote{ Updated version of the presentation at the {\it International Workshop on Cosmology \& Gravitational Physics,} December 15 - 16, 2005 Thessaloniki, Greece, {\it eds.} N. K. Spyrou, N. Stergioulas and C. Tsagas.}}

\author {\small \bf \underline{K. Kleidis} \thanks{kleidis@ihu.gr} \\ \small Department of Mechanical Engineering, International Hellenic University,\\
\small Serres Campus, 621.24 Serres, Greece}

\date{}

\maketitle

\vspace{0.5cm}

\begin{abstract}

The semiclassical interaction of the gravitational with a quantum scalar field is considered, in view of the renormalizability of the associated energy-momentum tensor in a n-dimensional curved spacetime resulting from a quad-ratic gravitational lagrangian. It is shown that, in this case, an additional coupling between the square curvature term, ${\cal R}^2$, and the quantum field needs to be introduced. The interaction so considered, discards any higher-order derivative terms from the associated gravitational field equations, but, at the expense, it introduces a geometric source term in the wave equation for the quantum field. Unlike the conformal coupling case, this term does not represent an additional mass; hence, in quadratic gravity theories, quantum fields can interact with gravity in a more generic way and not only through their mass (or energy) content.

\vspace{0.5cm}

Keywords: Quadratic Gravity - Semiclassical Theory - Quintessence

\end{abstract}

\vspace{1cm}

{\bf {1. Introduction}}\\

Nowadays, although there is a remarkable progress in understanding the quantum structure of the non-gravitational fundamental interactions, still there is no quantum framework consistent enough to describe gravity itself, leaving string theory as the most successful attempt towards that direction (Green et al. 1987, Schwarz 1999). In the context of General Relativity (GR), one usually resorts to a perturbative approach, in which string theory predicts corrections to the Einstein equations originating from higher-order curvature terms in the superstring action (Polchinsky 1998).

The mathematical background for {\em higher-order gravity theories} was formulated by Lovelock (1971). Accordingly, the most general gravitational Lagrangian consisting of terms up to the second order in the metric-tensor's derivatives reads \be {\cal L} = \sqrt {- g} \sum_{\nu=0}^{n/2} \lm_{\nu} {\cal L}^{(\nu)} \: , \ee where $\lm_{\nu}$ are constant coefficients, $n$ denotes the number of spacetime dimensions, $g$ is the determinant of the metric tensor, and ${\cal L}^{(\nu)}$ are functions of the Riemann curvature tensor, ${\cal R}_{ijkl}$, and its contractions, ${\cal R}_{ij}$ and ${\cal R}$, of the form \be {\cal L}^{(\nu)} = {1 \over 2^{\nu}} \dl_{i_1...i_{2{\nu}}}^{j_1...j_{2{\nu}}} {\cal R}_{j_1 j_2}^{i_1 i_2} ... {\cal R}_{j_{2{\nu}-1} j_{2{\nu}}}^{i_{2{\nu}-1} i_{2{\nu}}} \: , \ee where Latin indices refer to the n-dimensional spacetime and $\dl_{i_1...i_{2{\nu}}}^{j_1...j_{2{\nu}}}$ is the generalized Kronecker symbol. In Eq. (2), ${\cal L}^{(1)} = {1 \over 2} {\cal R}$ is the Einstein-Hilbert (EH) Lagrangian, while ${\cal L}^{(2)}$ is a particular combination of quadratic terms, known as the Gauss-Bonnett (GB) combination, since in four $(n = 4)$ dimensions it satisfies the functional relation \be {\dl \over \dl g^{i j }} \int \sqrt {- g} \left ( {\cal R}^2 - 4 {\cal R}_{i j}{\cal R}^{i j} + {\cal R}_{i j k l} {\cal R}^{i j k l} \right ) d^{\: 4} x = 0 \: , \ee representing the GB theorem (Kobayashi and Nomizu 1969). In view of Eq. (3), introduction of the GB term in the gravitational Lagrangian will not affect the four-dimensional field equations at all. However, in the context of the aforementioned superstring-induced perturbative approach, the most important contribution comes from the GB term (Mignemi and Stewart 1993). Since in four dimensions it is a total divergence, to render this term dynamical, one has to consider either a higher-dimensional background or to couple the GB term with a quintessential scalar field (or both).

The idea of a higher-dimensional spacetime has received much attention as a candidate for the unification of all fundamental interactions, including gravity, in the framework of {\em supergravity} and {\em superstrings} (Applequist et al. 1987, Green et al. 1987). In most higher-dimensional theories of gravity, at the present epoch the extra dimensions are assumed to form a compact manifold (the {\em internal space}) of very small size as compared to that of the three-dimensional visible space (the {\em external space}) and therefore they are unobservable at the energies currently available. This (so-called) {\em compactification} of the extra dimensions may be achieved, in a natural way, by adding a square-curvature term $({\cal R}_{ijkl} {\cal R}^{ijkl})$ in the EH action of the gravitational field
(M\"uller-Hoissen 1988). In this way, the higher-dimensional theories are closely related to those of non-linear gravitational Lagrangians and their combination probably yields a natural generalization of GR.

In the present paper, we explore this generalization with respect to the renormalizability of the energy-momentum tensor which acts as the source of gravity in the (semiclassical) interaction between the gravitational and a quantum matter field. In particular:

We briefly discuss how four-dimrnsional GR is modified by the introduction of a renormalizable energy-momentum tensor on the rhs of the Einstein field equations, first recognized by Calan et al. (1970). Using an analogous method, we explore the corresponding implications as regards a higher-dimensional quadratic gravity theory. We find that, in this case, the action functional describing the semiclassical interaction of a quantum scalar field with the classical gravitational one is modified and its variation with respect to the quantum field involved results in an inhomogeneous Klein-Gordon equation, the source term of which is purely geometric ($\sim {\cal R}^2$)!

\vspace{1.cm}

{\bf {2. Quadratic Interaction}}\\

Conventional gravity in $n$ dimensions implies that the dynamical behaviour of the gravitational field arises from an action principle involving the EH Lagrangian \be {\cal L}_{EH} = {1 \over 16 \pi G_n} \: {\cal R} \ee where $G_n = G V_{n-4}$ and $V_{n-4}$ denotes the volume of the internal space, formed by the extra spacelike dimensions. In this framework, we consider the semiclassical interaction between the gravitational and a massive quantum scalar field $\Phi (t, \vec{x})$ to the lowest order in $G_n$. The quantization of the field $\Phi (t, \vec{x})$ is performed by imposing canonical commutation relations on a hypersurface $t=$ constant (see, e.g., Isham 1981) \bea && \left [ \Ph (t, \vec{x}) \; , \; \Ph (t, {\vec{x}}^{\; \prime}) \right ] = 0 = \left [ \pi (t, \vec{x}) \; , \; \pi (t, {\vec{x}}^{\; \prime}) \right ] \nn \\ && \left [ \Ph (t, \vec{x}) \; , \; \pi (t, {\vec{x}}^{\; \prime}) \right ] = i \dl^{(n-1)} \left ( \vec{x} - {\vec{x}}^{\; \prime} \right ) \: , \eea where $\pi (t, \vec{x})$ is the momentum canonically conjugate to the field $\Ph (t, \vec{x})$. The equal-time commutation relations (5) guarantee the local character of the quantum field theory under consideration, thus attributing its time-evolution to the classical gravitational field equations (Birrell and Davies 1982).

In any local field theory the associated energy-momentum tensor is a very important object. Knowledge of its matrix elements is necessary to describe scattering in a relatively-weak external gravitational field. Therefore, in any quantum process in curved spacetime, it is admitted that the corresponding energy-momentum tensor is {\em renormalizable}; i.e., its matrix elements are {\em cut-off independent} (Birrell and Davies 1982). In this context, Calan et al. (1970) proved that the functional form of the renormalizable energy-momentum tensor involved in the semiclassical interaction between the gravitational and a quantum field in $n$ dimensions, is given by \be \Th_{ik} = T_{ik} - {1 \over 4} \: {n-2 \over n-1} \left [ {\Phi^2}_{; \: ik} - g_{ik} \Box \Phi^2 \right ] \: , \ee where the semicolon denotes covariant differentiation $(\nabla_k)$, $\Box = g^{ik} \nabla_i \nabla_k$ is the d' Alembert operator and \be T_{ik} = \Phi_{, i} \Phi_{, k} - g_{ik} {\cal L}_{mat} \ee is the {\em conventional} energy-momentum tensor of an (otherwise) free massive scalar field, with Lagrangian density of the form \be {\cal L}_{mat} = {1 \over 2} \left [ g^{ik} \Phi_{, i} \Phi_{, k} - m^2 \Phi^2 \right ] \: , \ee where comma denotes partial derivative. It is worth noting that the tensor (6) defines the same $n-$momentum and Lorentz generators as the conventional energy-momentum tensor. In fact, $\Th_{ik}$ can be obtained by an action principle involving \be S = \int \left [ f(\Phi) {\cal R} + {\cal L}_{mat} \right ] \sqrt{- g} d^n x \: , \ee where $f (\Phi)$ is an arbitrary, analytic function of $\Phi (t, \vec{x})$, the determination of which can be achieved by demanding that the rhs of the field equations resulting from Eq. (9) is given by Eq. (6). Accordingly, \be {\dl S \over \dl g^{ik}} = 0 \Rightarrow {\cal R}_{ik} - {1 \over 2} g_{ik} {\cal R} = - 8 \pi G_n \Th_{ik} = - {1 \over 2f} \left ( T_{ik} + 2 f_{; \: ik} - 2 g_{ik} \Box f \right ) \: , \ee which, to lowest order in $G_n$, results in \be f(\Phi) = {1 \over 16 \pi G_n} - {1 \over 8} {n-2 \over n-1} \Phi^2 \: . \ee In view of Eqs. (9) and (11), in any {\em linear gravity theory}, i.e., up to the first order in curvature tensor, renormalizability of the energy-momentum tensor associated to the interaction between a quantum scalar field and the classical gravitational one imposes that the appropriate coupling formula is determined through Hamilton's principle involving the action scalar \be S = \int \sqrt {-g} \left [ ( {1 \over 16 \pi G_n} - {1 \over 8} {n-2 \over n-1} \Phi^2 ) \: {\cal R} + {\cal L}_{mat} \right ] d^n x \: . \ee

On the other hand, both superstring theories (Candelas et al. 1985, Green et al. 1987) and the one-loop approximation of quantum gravity (Kleidis and Papadopoulos 1998), suggest that the presence of quadratic terms in the gravitational action is {\em a priori} expected.

In connection to the semiclassical interaction previously stated, we cannot help but wondering what the functional form of the corresponding renormalizable energy-momentum tensor would be if the simplest quadratic curvature term, ${\cal R}^2$, was included in the description of the classical gravitational field. To answer this question, by analogy to Eq. (9), we now consider the action principle \be {\dl \over \dl g^{ik}} \int \sqrt {-g} \left [ f_1 (\Ph) {\cal R} + \al f_2 (\Ph) {\cal R}^2 + {\cal L}_{mat} \right ] d^n x = 0 \: , \ee where both $f_1(\Ph)$ and $f_2(\Ph)$ are arbitrary, polynomial functions of $\Ph$. Eq. (13) yields \be {\cal R}_{ik} - {1 \over 2} g_{ik} {\cal R} = - {1 \over 2F} \left [ T_{ik} + 2 F_{; \: ik} - 2 g_{ik} \Box F + \al g_{ik} f_2 (\Ph) {\cal R}^2 \right ] \: , \ee where the function $F$ stands for the combination \be F = f_1 (\Ph) + 2 \al {\cal R} f_2 (\Ph) \: . \ee For $\al = 0$ and to lowest order in $G_n$ (but to every order in the coupling constants of the quantum field involved), we must have \be {\cal R}_{ik} - {1 \over 2} g_{ik} {\cal R} = - 8 \pi G_n \Th_{ik} \: ,  \ee where, $\Th_{ik}$ is given by Eq. (6). Accordingly, we obtain $f_1 (\Ph) = f(\Ph)$, i.e., a function quadratic in $\Ph$ [cf. Eq. (11)]. As regards Eq. (14), on dimensional grounds we expect that \be F \sim \Ph^2 \ee and, therefore, $\al {\cal R} f_2(\Ph) \sim \Ph^2$, as well. However, we already know that $[{\cal R}] \sim [\Ph]$ (see, e.g., Whitt 1984) and, therefore, $f_2 (\Ph) \sim \Ph$. {\footnote {In fact, $[{\cal R}] \sim [\Phi]^{4 \over n-2}$ and, therefore, $f_2 \sim \Phi^{2 {n-4 \over n-2}}$. Hence, in order to render the coupling constant $\al$ dimensionless, one should consider $n=6$, i.e., $f_2 \sim \Phi$ only in six dimensions.}} In particular, \be F(\Phi) = {1 \over 16 \pi G_n} - {1 \over 8} {n-2 \over n-1} \Ph^2 + 2 \al {\cal R} \Ph \: . \ee In Eq. (18), the coupling parameter $\al$ incorporates any arbitrary constant that can be introduced in the definition of $f_2(\Ph)$. Accordingly, the action describing the semiclassical interaction of a quantum scalar field with the classical gravitational one in a {\em quadratic gravity theory}, i.e., up to the second order in curvature tensor, is written in the form \be S = \int \sqrt {-g} \left [ ( {1 \over 16 \pi G_n} - {1 \over 2} \ks_n \Ph^2 ) {\cal R} + \al {\cal R}^2 \Ph + {\cal L}_{mat} \right ] d^n x \: , \ee where \be \ks_n = {1 \over 4} {n-2 \over n-1} \ee is the so-called {\em conformal coupling} parameter (Birrell and Davies 1982). In this case, the associated gravitational field equations (14) result in \be {\cal R}_{ik} - {1 \over 2} g_{ik} {\cal R} = -8 \pi G_n \: (\Th_{ik} + \al S_{ik}) \: , \ee where \be S_{ik} = g_{ik} \: {\cal R}^2 \Ph \: . \ee Clearly, the rhs of Eq. (21) represents the renormalizable energy-momentum tensor in a quadratic theory of gravity. Notice that, as long as $\al \neq 0$, this tensor contains the extra {\em source} term $S_{ik}$. In spite of the presence of this term, the generalized energy-momentum tensor still remains renormalizable. This is due to the fact that, the set of quantum operators $\lbrace \Phi, \Phi^2, \Box \Phi \rbrace$ is closed under renormalization, as it can be verified by straightforward power counting (see, e.g., Callan et al. 1970).

Eq. (22) implies that, in the semiclassical interaction of the gravitational with a quantum scalar field in more than four dimensions, the quadratic curvature term (i.e., pure {\em global} gravity) acts as a {\em source} of the quantum field. Indeed, variation of Eq. (19) with respect to $\Ph (t, \vec{x})$ leads to the following equation of propagation \be \Box \Ph + m^2 \Ph + \ks_n {\cal R} \Ph = \al {\cal R}^2 \: , \ee i.e., an inhomogeneous Klein-Gordon equation in curved spacetime. It is worth pointing out that, in Eq. (19) and/or Eq. (23), the coupling constant $\al$ remains dimensionless (this is also the case for the corresponding action) only as long as \be n = 6 \: , \ee indicating the appropriate spacetime dimensions for this quadratic semiclassical theory to hold, without introducing any additional arbitrary length scales.

\vspace{.5cm}

{\bf {3. Summary and Conclusions}}\\

{\em Summarizing}, a novel coupling between the square curvature term, ${\cal R}^2$, and the quantum scalar field, $\Ph (t, \vec{x})$, should be introduced in order to yield the correct renormalizable energy-momentum tensor in a higher-dimensional, semiclassical quadratic gravity theory. This coupling discards any higher-order derivative terms from the gravitational field equations, but it introduces an unexpected {\em geometric source} in the wave equation for the quantum field. In this case, unlike the conventional conformal coupling $(\sim {\cal R} \Ph^2)$, the quantum field interacts with gravity not only through its mass (or energy) content $(\sim \Ph^2)$, but, also, in a more generic (global) way $({\cal R}^2 \Ph)$.

\vspace{.5cm}

{\bf Acknowledgements} \\ This research was supported by the Greek Ministry of Education, through the PYTHAGORAS program.

\vspace{1.cm}

{\bf {References}}\\

\rfnce Applequist T., Chodos A., and Freund P., 1987, {\em Modern Kaluza-Klein Theories}, Addison-Wesley, Menlo Park, CA.

\rfnce Birrell N. D. and Davies P. C. W., 1982, {\em Quantum Fields in Curved Space}, Cambridge University Press, Cambridge.

\rfnce Calan C. G., Coleman S., and Jackiw R., 1970, Ann. Phys. {\bf 59}, 42.

\rfnce Candelas P., Horowitz G. T., Strominger A., and Witten E., 1985, Nucl. Phys. {\bf B 258}, 46.

\rfnce Green M. B., Schwartz J. H., and Witten E., 1987, {\em Superstring Theory}, Cambridge University Press, Cambridge.

\rfnce Isham C. J., 1981, {\it Quantum Gravity - An overview}, in {\em Quantum Gravity: a Second Oxford Symposium}, C. J. Isham, R. Penrose and D. W. Sciama (eds.), Clarendon, Oxford.

\rfnce Kleidis K. and Papadopoulos D. B., 1998, Class. Quantum Grav. {\bf 15}, 2217.

\rfnce Kobayashi S. and Nomizu K., 1969, {\em Foundations of Differential Geometry II}, Wiley InterScience, NY.

\rfnce Lovelock D., 1971, J. Math Phys. {\bf 12}, 498.

\rfnce Mignemi S. and Stewart N. R., 1993, Phys. Rev. {\bf D 47}, 5259.

\rfnce M\"uller-Hoissen F., 1988, Class. Quantum Grav. {\bf 5}, L35.

\rfnce Polchinski J. 1998, {\em String Theory}, Cambridge University Press, Cambridge.

\rfnce Schwartz J. H., 1999, Phys. Rep. {\bf 315}, 107.

\rfnce Whitt B., 1984, Phys. Lett. {\bf B 145}, 176.

\end{document}